# WiEps: Measurement of Dielectric Property with Commodity WiFi Device – An application to Ethanol/Water Mixture

Hang Song, *Member, IEEE*, Bo Wei, *Member, IEEE*, Qun Yu, Xia Xiao, *Member, IEEE*, and Takamaro Kikkawa, *Fellow, IEEE*

*Abstract*—WiFi signal has become accessible everywhere, providing high-speed data transmission experience. Besides the communication service, channel state information (CSI) of the WiFi signals is widely employed for numerous Internet of Things (IoT) applications. Recently, most of these applications are based on analysis of the microwave reflections caused by physical movement of the objective. In this paper, a novel contactless wireless sensing technique named WiEps is developed to measure the dielectric properties of the material, exploiting the transmission characteristics of the WiFi signals. In WiEps, the material under test is placed between the transmitter antenna and receiver antenna. A theoretical model is proposed to quantitatively describe the relationship between CSI data and dielectric properties of the material.

During the experiment, the phase and amplitude of the transmitted WiFi signals are extracted from the measured CSI data. The parameters of the theoretical model are calculated using measured data from the known materials. Then, WiEps is utilized to estimate the dielectric properties of unknown materials. The proposed technique is first applied to the ethanol/water mixtures. Then, additional liquids are measured for further verification. The estimated permittivities and conductivities show good agreement with the actual values, with the average error of 4.0% and 8.9%, respectively, indicating the efficacy of WiEps. By measuring the dielectric property, this technique is promising to be applied to new IoT applications using ubiquitous WiFi signals, such as food engineering, material manufacturing process monitoring, and security check.

## I. INTRODUCTION

RECENT years, with the rapid evolution of Internet of Things (IoT), wireless sensing using ubiquitous signals has been widely studied [1-4]. Compared with vision-based or sensor-based approaches, wireless sensing does not require any additional sensors deployed on human body and it can operate without light. Therefore, it has gained extensive research attention worldwide. Owing to the fine-grained channel state information (CSI) [5], numerous human-centric applications have been proposed using off-the-shelf WiFi devices. Gu *et al.* developed MoSense, a motion detection system which can identify human presence and activities [6]. Liu *et al.* proposed a sleep monitoring system which can recognize the postures and vital signs [7]. Zhang *et al.* developed BreathTrack, a system which can be used to accurately track the human breath status indoor [8]. The emerging artificial intelligence techniques are also incorporated with the WiFi signals to improve the activity recognition accuracy [9, 10]. Wang *et al.* proposed the LiFS, a submeter position system which is low human-effort and device-free [11]. WiMorse, a Morse-code input system is proposed for people who have disease by tracking the finger movements [12]. Besides, other systems are proposed for specific use such as emotion detection [13], non-intrusive user identification [14], and vehicle speed estimation [15].

Although a lot of systems have been developed for different purposes with WiFi signals as mentioned above, the physical basis of these applications is similar. The variations of the signal amplitude and phase caused by *physical movement* are utilized. According to electromagnetic wave theory [16], when the emitted microwave encounters an interface with a different dielectric property, the wave will be partially reflected and partially transmitted through. The variations caused by physical movement is related to reflected signals. Therefore, the majority of recent researches are concentrated on the reflection part of the microwave. In fact, the dielectric characteristics will also influence the transmission of the microwave signal, resulting in the change of signal amplitude and phase. There are a few works utilizing the change of signal caused by dielectric property for wheat moisture detection [17], fire event detection [18], and material classification [19]. However, these applications are mainly formulated as classification problems. The core techniques are feature extraction and classification methods. While the physical basis of these applications, which is *dielectric property change*, is not revealed and thoroughly

This work was supported in part by the JSPS KAKENHI Grant Number 20K14740 and in part by NSFC Grant Number 61271323. (*Corresponding author: Bo Wei and Hang Song*)

Hang Song, Qun Yu, and Xia Xiao are with Tianjin Key Laboratory of Imaging and Sensing Microelectronic Technology, School of Microelectronics, Tianjin University, Tianjin 300072, China. (e-mail: songhang168@tju.edu.cn)

Bo Wei is with Department of Computer Science and Communication Engineering, Waseda University, 3-4-1 Okubo, Shinjuku-ku, Tokyo 169-8555, Japan. (e-mail: weibo0504@fuji.waseda.jp).

Takamaro Kikkawa is with Research Institute for Nanodevice and Bio Systems, Hiroshima University, Hiroshima 739-8527, Japan.

analyzed. In this paper, the transmitted WiFi signal through a material is theoretically analyzed and it is utilized to quantitatively measure the dielectric properties of the materials.

Measuring the dielectric property is essential in many applications since it reflects the physical and chemical characteristics of the materials [20]. When the components of a certain material changes, the permittivity and conductivity will be influenced. Therefore, by knowing the dielectric properties of the materials, the constituents can be analyzed. Dielectric characterization has been widely applied to a variety of areas such as monitoring the quality of the composite material during manufacturing [21], food engineering [22-24], analyzing the material chemical features [25], security check [26], and biomedical engineering [27-30]. However, most of current measurement methods require the use of expensive commercial equipment such as vector network analyzer (VNA). This limits the deployment in wider scenarios for small business or individual user at a low cost.

In this work, a new system named WiEps is developed for dielectric characterization with commodity WiFi devices. *Eps* stands for *epsilon* which is commonly utilized to note the permittivity. In WiEps system, two computers equipped with the Intel 5300 NICs are utilized to transmit and receive signal, respectively. On the transmitter side, one antenna is connected. While on the receiver side, two antennas are employed to receive the signal. The material to be tested is placed on the line-of-sight (LOS) between the transmitter and one of the receivers. The antennas used are also commercial goods for WiFi communication system without any modification. During the experiment, the signals are emitted towards the material under test (MUT) and the transmitted signals are captured by the receiver antennas. CSI data are recorded by the NIC and then processed to extract the amplitude and phase information of the received signal. A theoretical model is proposed to quantitatively describe the relationship between transmitted signal characteristics and dielectric properties of the material. In this model, the influence of the multipath effect is incorporated.

For evaluation, WiEps is first applied to measure the ethanol/water mixture. The ethanol/water solutions with different concentrations are prepared and put into a container in turn. Then, the signal transmission experiments are conducted. In order to calculate the parameters of the theoretical model, the dielectric properties of the solutions are also measured using standard open-ended coaxial probe method with VNA. For further validation, additional liquids are measured using WiEps. The estimated dielectric properties agree well with the actual values, with the average error of 4.0% and 8.9% across all materials for dielectric constant and conductivity, respectively. The results demonstrate that the proposed WiEps system is effective in measuring both the dielectric constant and conductivity of the material, and it is promising to be applied to new IoT applications using ubiquitous WiFi signals.

The main contributions of this work are as follows:
1) We investigate the transmitted signals through the material and analyze the influence of the dielectric properties on the channel state response.

2) We propose WiEps, a dielectric characterization system which can estimate the dielectric property of a material with ambient WiFi signals. To the best of our knowledge, this is the first work which theoretically analyze the transmission characteristics of the WiFi signals and utilize the CSI data for dielectric property measurement.

3) We propose a theoretical model which quantitatively describes the relationship between the dielectric properties and the transmitted WiFi signals. In this model, the influence of multipath effect is taken into consideration. Therefore, it can work under indoor condition without any specific requirement.

4) We implement the WiEps using commodity WiFi device and apply it to measure the dielectric properties of ethanol/water mixtures and other liquids. The experiment results demonstrate the efficacy of the proposed system.

The rest of the paper is organized as follow. Section II briefly review the methods for dielectric characterization. Section III depicts the detailed system design of WiEps. Section IV illustrates the theoretical model for measuring dielectric property using WiFi signals. Section V presents the experimental evaluation of the proposed system with ethanol/water mixtures and other liquids. Finally, the conclusion is given in Section VI.

## II. RELATED WORK

Generally, there are two kinds of approaches to measure the dielectric properties, the resonant and non-resonant methods [31]. In the resonant method, customized cavity or resonator is designed and perturbation principle is utilized to analyze the resonant shift with and without the existence of MUT. Rajab *et al.* proposed a dielectric characterization technique using a resonant nonradiative dielectric waveguide structure [32]. Zhu *et al.* developed a hollow coaxial cable Fabry–Perot resonator for measuring liquid dielectric constant and applied it to monitor the evaporation of ethanol in ethanol/water mixtures [33]. Koirala *et al.* measured the water/ethanol solution using a spiral-coupled passive micro-resonator sensor [34]. Su *et al.* developed a flexible complementary spiral resonator and attached it to robotic hand for measuring the objects grasped [35]. Other devices such as metamaterial-based sensors [36, 37] and microfluidic sensor [38] are developed for measuring the dielectric properties of chemicals and liquids.

In the non-resonant method, customized waveguide or transmission lines are fabricated and the lumped models are utilized to analyze the impedance change with the presence of MUT. Bois *et al.* developed a plug-loaded two-port waveguide measurement technique for dielectric characterization of granular and liquid materials [21]. Bobrov *et al.* proposed a technique for measuring the dielectric property of soil in a wideband frequency using a single coaxial cell [39]. Pantoja *et al.* proposed a new model to estimate the complex permittivity of the soil using the two-port coaxial probe [40]. Meaney *et al.* developed a transmission-based open-ended coaxial probe for dielectric characterization in clinical use [41].

Besides the conventional methods which are based on VNA, there are other methods developed for dielectric measurement.

Guo *et al.* proposed an improved device based on parallel-plate capacitance method for measuring the rocks with the impedance analyzer or VNA [42]. Gutierrez *et al.* developed a field-deployable system to measure the complex permittivity of the improvised explosives using M-sequence device [26]. Bertling *et al.* measured the ethanol content of liquid solution using the terahertz quantum cascade laser [43]. Dhekne *et al.* developed a system using UWB signals to measure the permittivities of liquids [44].

## III. System Design

The system design and configuration of WiEps is shown in Fig. 1. In WiEps system, two commodity WiFi devices, Intel 5300 NICs are installed on two separate personal computers. Three antennas are utilized to emit and receive signals. One emitter antenna Tx is connected to one NIC. The other two antennas Rx1 and Rx2 are connected to the other NIC and utilized as the receivers. The material to be measured is placed in between the Tx and Rx2 on the LOS. The thickness of the material is denoted as *d*. In this study, a acrylics-made container is utilized to hold the water/ethanol mixture and other materials. Then, the thickness of the gap which contains the material is *d*. The container is originally used for storing papers and it is sold on market. The antennas are also commercial goods for WiFi communication system without any modification.

During the experiment, the communication is conducted between two computers. The modulated signal is emitted from Tx and received by Rx1 and Rx2. As shown in Fig. 1, the red dash lines represent the wave propagation on LOS where the effect of the material is included. Since the experiment is assumed to be carried out in indoor environment, there are multipath effects which influence the received signals by Rx antennas. The multipath waves are illustrated as purple dash lines in Fig. 1. After being captured, the received signals are processed by NIC and CSI data are recorded.

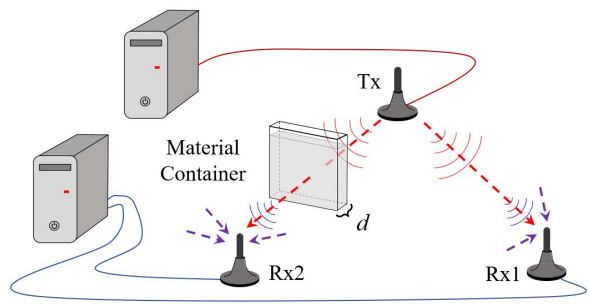

Fig. 1. Schematic diagram of the proposed WiEps system. (Red and purple dash lines represent the wave propagation in LOS and multipath effect, respectively.)

WiFi standard 802.11n utilized the orthogonal frequency division multiplexing (OFDM) technology for communication [45]. The recorded CSI reveals the channel response features of the subcarriers. Specifically, the utilized device measures the CSI data of a subgroup of 30 within the 64 subcarriers. When there is a material presenting at the propagation path in between Tx and Rx2, the received signal by Rx2 will be influenced and the channel state will change accordingly. Meanwhile, a consistent reference channel is utilized to synchronize the data. In order to quantitatively characterize the relationship between transmitted signal characteristics and dielectric properties of the material, a theoretical model is proposed. In the model, the influence of the multipath effect is incorporated. Therefore, there is no special requirement of the environment for deploying the WiEps system.

## IV. Theoretical Model

Define the received signal by the antenna Rx2 in complex form as $A_{r2} \cdot e^{j\theta_{r2}}$, where $A_{r2}$ and $\theta_{r2}$ are the amplitude and phase of signal. Since the received signal is the superposition of multipath components, it can be expressed as follows:

$$A_{r2} \cdot e^{j\theta_{r2}} = A_{LOS} \cdot e^{j\theta_{LOS}} + A_m \cdot e^{j\theta_m} \qquad (1)$$

where $A_{LOS}$ and $\theta_{LOS}$ are the amplitude and phase of signal transmitted on LOS where the material exists. While $A_m \cdot e^{j\theta_m}$ represents the total effect of other multipath signals. When the microwave propagates in a dissipative medium, the amplitude is attenuated by a factor of $e^{-k_I \cdot d}$. Meanwhile, the phase change in the medium is calculated as $-k_R \cdot d$ [16]. *d* is the propagating distance which is equivalent to the thickness of the material as shown in Fig. 1. $k_R$ and $k_I$ are the coefficients calculated as follows:

$$k_R = \omega\sqrt{\mu_0 \varepsilon_0 \varepsilon_r}\left[\frac{1}{2}\left(\sqrt{1 + \frac{\sigma^2}{(\varepsilon_0 \varepsilon_r \omega)^2}} + 1\right)\right]^{1/2} \qquad (2)$$

$$k_I = \omega\sqrt{\mu_0 \varepsilon_0 \varepsilon_r}\left[\frac{1}{2}\left(\sqrt{1 + \frac{\sigma^2}{(\varepsilon_0 \varepsilon_r \omega)^2}} - 1\right)\right]^{1/2} \qquad (3)$$

where $\omega$ is the angular frequency of the wave, $\mu_0$ and $\varepsilon_0$ are the magnetic permeability and dielectric permittivity of vacuum, $\varepsilon_r$ and $\sigma$ are the relative dielectric constant and conductivity of the medium. Considering the attenuation effect of the material, (1) can be rewritten as:

$$A_{r2} \cdot e^{j\theta_{r2}} = A'_{LOS} \cdot e^{-k_I \cdot d} \cdot e^{j(\theta'_{LOS} - k_R \cdot d)} + A_m \cdot e^{j\theta_m} \quad (4)$$

where the $A'_{LOS} \cdot e^{j\theta'_{LOS}}$ implies the signal without the effect of the material.

The received signals can also be expressed as the multiplication of transmitted signal using the channel frequency response in frequency domain:

$$R(f) = S(f) \cdot H(f) \qquad (5)$$

where $S(f)$ and $R(f)$ are the spectrums of the transmitted and received signal. $H(f)$ is the channel response which is measured by the NIC and revealed as the CSI data. Then, the signals $A_{r2} \cdot e^{j\theta_{r2}}, A_m \cdot e^{j\theta_m}, A'_{LOS} \cdot e^{j\theta'_{LOS}}$ can be represented as:

$$A_{r2} \cdot e^{j\theta_{r2}} = S(f) \cdot \|H_{r2}(f)\| \cdot e^{j\theta_{Hr2}} \tag{6}$$

$$A_m \cdot e^{j\theta_m} = S(f) \cdot \|H_m(f)\| \cdot e^{j\theta_{Hm}} \tag{7}$$

$$A'_{LOS} \cdot e^{j\theta'_{LOS}} = S(f) \cdot \|H_l(f)\| \cdot e^{j\theta_{Hl}} \tag{8}$$

where $H_{r2}(f)$, $H_m(f)$, and $H_l(f)$ are the channel responses corresponding to the receiver Rx2, multipath, and LOS without material, respectively. Insert (6)~(8) into (4) and divide out $S(f)$, the following equation can be obtained:

$$\|H_{r2}(f)\| \cdot e^{j\theta_{Hr2}} = \|H_l(f)\| \cdot e^{j\theta_{Hl}} \cdot e^{-k_I \cdot d - j \cdot k_R \cdot d} + \|H_m(f)\| \cdot e^{j\theta_{Hm}} \tag{9}$$

$H_{r2}(f)$ in (9) can be extracted from the measured CSI data. While $H_m(f)$, and $H_l(f)$ cannot be measured directly and these parameters vary in different environment. Therefore, before the measurement of unknown material, the system needs a calibration process using materials whose dielectric property are already available.

In a configured system, the channel response is expected to be consistent in a steady condition. However, in this wireless communication system, the clocks of the transmitter and the receiver are not synchronized. Therefore, the phase of the measured CSI such as $\theta_{Hr2}$ is changing randomly even when measuring the same material. This kind of raw CSI data from the WiFi device cannot reveal the influence from the material. Nevertheless, the sampling clocks of the different antennas on the receiver side are the same. Figure 2 shows an illustration of the recorded channel response data in two measurements of different materials. It can be observed that the phase information of $H_{r1}$ in two cases varies a lot, while it should be the same since the channel between Tx and Rx1 is not changed. Therefore, the comparison of the absolute phase information of $H_{r2}$ cannot reveal the phase shift value caused by the material. To overcome this problem, all the channel response data are multiplied by a factor $e^{-j\theta_{Hr1}}$, which is equivalent to adjusting the phases of the data to make the phase of $H_{r1}$ be zero as shown in Fig. 2(c). After the phase adjustment, all the phases are synchronized to $H_{r1}$. Then, the variation of differential phase $\Delta\theta_{r2} = \theta_{Hr2} - \theta_{Hr1}$ implies the effect of the material. Denote the data after adjustment as phase adjusted CSI data.

Multiply both sides of (9) by $e^{-j\theta_{Hr1}}$, the following equation can be obtained:

$$\|H_{r2}(f)\| \cdot e^{j\theta_{Hr2}} \cdot e^{-j\theta_{Hr1}} = \|H_l(f)\| \cdot e^{j\theta_{Hl}} \cdot e^{-j\theta_{Hr1}} \cdot e^{-k_I \cdot d - j \cdot k_R \cdot d} + \|H_m(f)\| \cdot e^{j\theta_{Hm}} \cdot e^{-j\theta_{Hr1}} \tag{10}$$

Introduce three variables $\Delta\theta_{Hr2} = \theta_{Hr2} - \theta_{Hr1}$, $\Delta\theta_{Hl} = \theta_{Hl} - \theta_{Hr1}$, and $\Delta\theta_{Hm} = \theta_{Hm} - \theta_{Hr1}$. Insert the $\Delta\theta_{Hr2}$, $\Delta\theta_{Hl}$, and $\Delta\theta_{Hm}$ into (10), the equation can be rewritten as follows:

$$\|H_{r2}(f)\| \cdot e^{j\Delta\theta_{Hr2}} = \|H_l(f)\| \cdot e^{j\Delta\theta_{Hl}} \cdot e^{-k_I \cdot d - j \cdot k_R \cdot d} + \|H_m(f)\| \cdot e^{j\Delta\theta_{Hm}} \tag{11}$$

By using (11), the relationship between the material dielectric property and the measured CSI data is characterized with two complex coefficients to be determined, which are $\|H_l(f)\| \cdot e^{j\Delta\theta_{Hl}}$ and $\|H_m(f)\| \cdot e^{j\Delta\theta_{Hm}}$. The effects from other objects' reflected signals are included in $\|H_m(f)\| \cdot e^{j\Delta\theta_{Hm}}$. Since other objects are kept stationary during the whole experiment, $\|H_m(f)\| \cdot e^{j\Delta\theta_{Hm}}$ is considered to be consistent. By using the materials whose dielectric properties are already known, the coefficients can be calculated.

Rewrite (11) as the following:

$$e^{-k_I \cdot d - j \cdot k_R \cdot d} = \frac{\|H_{r2}(f)\| \cdot e^{j\Delta\theta_{Hr2}} - \|H_m(f)\| \cdot e^{j\Delta\theta_{Hm}}}{\|H_l(f)\| \cdot e^{j\Delta\theta_{Hl}}} \tag{12}$$

Denote the result of the right side of (12) as $B \cdot e^{j\theta_B}$. Then, the following can be obtained:

$$k_I = -\frac{\ln B}{d} \tag{13}$$

$$k_R = -\frac{\theta_B}{d} \tag{14}$$

Divide $k_R$ by $k_I$ using (3) and (2), the following equation can be obtained:

$$\left(\frac{k_R}{k_I}\right)^2 = \frac{\sqrt{1 + \frac{\sigma^2}{(\varepsilon_0 \varepsilon_r \omega)^2}} + 1}{\sqrt{1 + \frac{\sigma^2}{(\varepsilon_0 \varepsilon_r \omega)^2}} - 1} \tag{15}$$

Introduce $n = \sigma/\varepsilon_r$ to (15), thus (15) can be rewritten as:

$$\left[\left(\frac{k_R}{k_I}\right)^2 - 1\right] \cdot \sqrt{1 + \frac{n^2}{(\varepsilon_0 \omega)^2}} = 1 + \left(\frac{k_R}{k_I}\right)^2 \tag{16}$$

By using (16), $n$ can be calculated as:

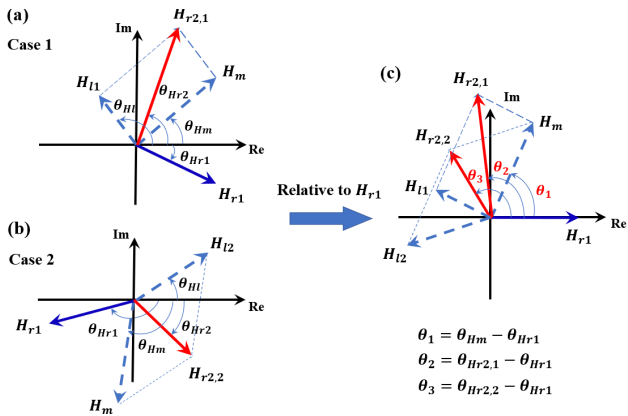

Fig. 2. Illustration of phase information revealed by the channel response data. (a)(b) Two independent measurement cases. (c) Adjust the phase relative to that of the reference channel $H_{r1}$.

$$n = \varepsilon_0 \omega \sqrt{\left[\frac{1 + \left(\frac{k_R}{k_I}\right)^2}{\left(\frac{k_R}{k_I}\right)^2 - 1}\right]^2 - 1} \qquad (17)$$

Substitute $\sigma/\varepsilon_r$ with $n$ in (2), the relative dielectric constant can be calculated as:

$$\varepsilon_r = \frac{2k_R^2}{\omega^2 \mu_0 \varepsilon_0 \left(\sqrt{1 + \frac{n^2}{\omega^2 \varepsilon_0^2}} + 1\right)} \qquad (18)$$

Subsequently, the conductivity can be obtained by:

$$\sigma = n \cdot \varepsilon_r \qquad (19)$$

With the derivation above, the dielectric constant $\varepsilon_r$ and conductivity $\sigma$ of a material can be obtained from the measured channel response data.

Figure 3 summarizes the flowchart of WiEps system for the dielectric characterization. The operation is divided into two stages, which are the calibration stage and estimation stage. In the calibration stage, the measurements are conducted using materials with known dielectric properties. Then, the coefficients which determine current WiEps system are calculated using the phase adjusted CSI data and the factors $k_I$ and $k_R$ as described in (11). In the estimation stage, the measurements are carried out using the materials whose dielectric properties are unknown. Subsequently, the factors $k_I$ and $k_R$ are calculated using the phase adjusted CSI data and the coefficients as depicted in (12)~(14). Finally, the dielectric constant and conductivity of the material are estimated as described in (17)~(19).

## V. EXPERIMENT EVALUATION

The performance of the proposed WiEps system is first evaluated by applying to the measurement of ethanol/water mixtures. In the evaluation, the solutions with different ABVs are prepared. The ABVs are from 0% to 90% at a step of 10%. The Chinese liquor (Baijiu) with the ABV of 46% and 56% are utilized to assess the system. Besides, additional liquids are tested for further validation of the system.

### A. Dielectric Property Measurement with Standard Technique

In order to calibrate the WiEps system and evaluate the correctness of the results by the proposed method, the dielectric properties of the materials are measured in advance by standard technique with VNA and open-ended coaxial probe [27-29, 46]. Figure 4 shows the measurement setup. The ROHDE & SCHWARZ ZVA40 VNA with the frequency range of 10 MHz ~ 40 GHz and SPEAG DAK-3.5 probe with the frequency range of 200 MHz ~ 20 GHz are utilized. The cables and the probe are fixed carefully. They are not touched through the whole process. The calibration is carried out before measurement using Open/Short/Load (OSL) technique. The ethanol/water solutions with different ABVs and other materials are stored in separate bottles. During the measurement, the cushion is first moved away and the probe is immersed into the solution by moving the bottle. The probe is kept static and it is ensured that there are no air bubbles attaching on the probe. Then, the cushion is set under the bottle to keep the measurement system stable. After the measurement, the cushion and bottle are moved away and the head of the probe is dried by tissue. The bottle is changed in turn and the measurement is repeated for all materials. The measured data are sent to the PC connected to the VNA and processed by the DAK software to get the dielectric constant and conductivity values.

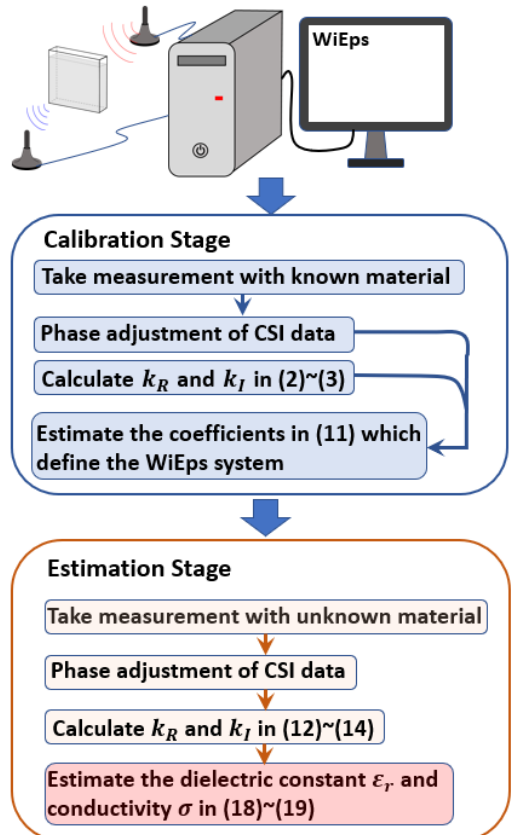

Fig. 3. Flowchart of the measurement of dielectric property using WiEps system.

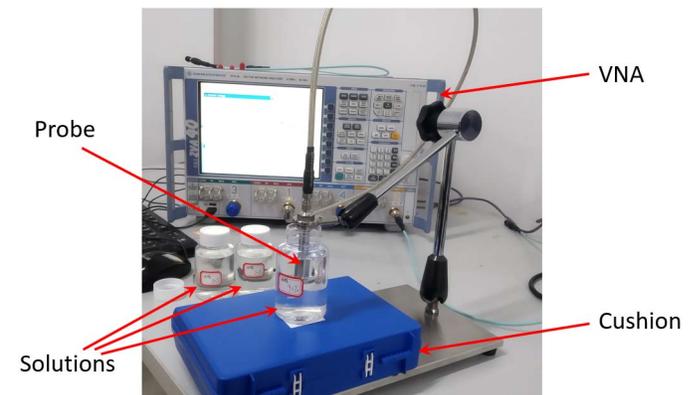

Fig. 4. Dielectric property measurement setup for the ethanol/water solutions and other materials using VNA and open-ended coaxial probe.

The dielectric properties measured in a wide frequency range by the standard technique are shown in Fig. 5. It can be observed that the dielectric constant decreases with the frequency increasing. While the conductivity is increasing when the frequency increases. The conductivity implies the dissipation effect of electromagnetic energy in the materials. The increasing conductivity indicates that the energy of the microwave is absorbed more at higher frequencies. Regarding to different ABVs, the dielectric constant increases monotonically with the reduction of alcohol volume. While the conductivity does not show such monotonic behavior with the change of ABV. The relationship between the ABV rate and wideband dielectric properties has been studied mathematically in [46, 47] using dispersive models. While in this work, the frequency range of the microwave is narrow. Therefore, the dielectric properties at a single frequency is considered. The dash line shows the values at 5.32 GHz, which is the chosen operation frequency of WiEps system in the evaluation. The dielectric properties at 5.32 GHz will be utilized for calibrating WiEps and comparing the results.

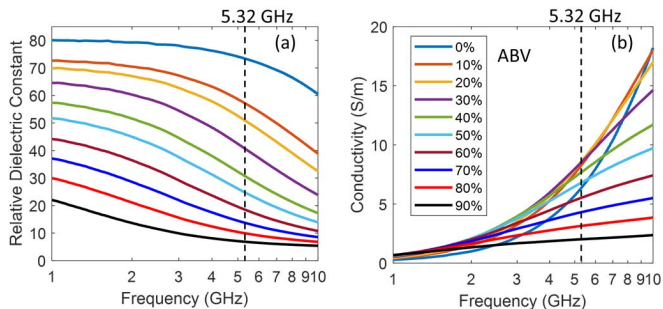

Fig. 5. Dielectric properties of the solutions with different ABVs measured by standard technique. (a) Relative dielectric constant. (b) Conductivity.

### B. Measurement with WiEps System

The experiment configuration of WiEps system in operation is shown in Fig. 6. The antennas and the container are fixed on the table. Tx1 is connected to port A of the NIC on transmitter side. Rx1 and Rx2 are connected to port A and B of the NIC on receiver side. The size of the container is A5 standard with the length of 210 mm and the height of 148 mm. The gap of the container is 2 mm. The communication channel 64 with the center frequency of 5.32 GHz and bandwidth of 20 MHz is chosen in the experiment [45]. Tx1 is set to transmit signals every 0.05 s and each measurement lasts 20 s. It should be noted that the physical movement of human affects the received signals significantly. Therefore, during the experiment, after loading the material, the operator starts the measurement and gets away from the system immediately to avoid the influence of human on the final experiment results.

After measurement, both the coarse-grained received signal strength indicator (RSSI) and the fine-grained CSI data are recorded. RSSI indicates the whole received power by the receiver antenna and CSI reveals the channel response by each subcarrier. However, the values of RSSI and CSI are relative to some internal references. The raw data cannot be utilized for quantitative calculation. In the theoretical derivation, the amplitude of the channel response such as $\|H_{r2}(f)\|$ in (11) is expected to be in voltage. Therefore, the CSI data need pre-processing to re-scale the unit. The total received power can be calculated by:

$$P_{total} = 10\left\{\frac{10\log_{10}\left[10^{\left(\frac{RSSIa}{10}\right)}+10^{\left(\frac{RSSIb}{10}\right)}+10^{\left(\frac{RSSIc}{10}\right)}\right]-AGC-C}{10}\right\} \quad (20)$$

where $RSSIa$, $RSSIb$, and $RSSIc$ are the recorded RSSI at three antenna ports, respectively. $AGC$ is the automatic gain control factor which is implemented in the hardware to amplify the received signal adaptively before sampling. $C$ is a constant which implies the internal reference. Subsequently, a rescale factor can be obtained by:

$$\alpha = \frac{P_{total}}{\left(\sum CSIa \cdot CSIa^H + \sum CSIb \cdot CSIb^H + \sum CSIc \cdot CSIc^H\right)/30} \quad (21)$$

where $CSIa$, $CSIb$, and $CSIc$ are the CSI data recorded at three ports, respectively. $CSIa^H$, $CSIb^H$, $CSIc^H$ are the corresponding conjugates. Then, all the CSI data are multiplied by $\sqrt{\alpha}$ and the units are in voltage.

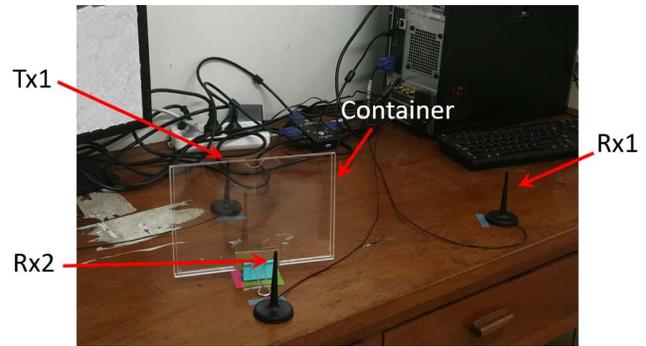

Fig. 6. Experiment configuration of WiEps system.

Figure 7 shows an example of the amplitude $\|H_{r2}(f)\|$ and adjusted phase $\Delta\theta_{Hr2}$ for all recorded 30 subcarriers at Rx2 in time dimension. It can be observed that there are large variations at the first 5 seconds (red box in Fig. 7) both in phase and amplitude. This is caused by the human activity where the operator leaves from the system after measurement starts. The signals become relatively stable after 5 seconds. As for a single carrier, there are still some random noise in the data. The time-averaging method is used to mitigate the noise. To ensure that the system is completely stable and the influence of human is avoided, the averaging range is chosen from 10 s to 20 s. The averaged values are employed for calculation of the dielectric property.

Figure 8 shows the averaged amplitudes and adjusted phases of all the ethanol/water mixtures with respect to subcarrier number. Generally, microwave signals propagate slower in materials with larger dielectric constant. Thus, the phase will be lagged more. It can be observed that with the increase of ABV, the phase become larger at all subcarrier frequencies as shown in Fig. 8(a). The results coincide with the dielectric properties measured in Section V.A. As for the amplitude, the microwave attenuates more quickly in materials with higher conductivity. However, the values for 0% and 90% at higher frequencies tend to be similar while the conductivities of the two ABVs are quite different. This is caused by the multipath effect. Meanwhile, the difference of amplitude is relatively smaller for various ABVs and no monotonic relationship is found between the amplitude and ABV. In order to quantitatively characterize the relationship between the CSI data and dielectric property, the model proposed in Section IV is employed.

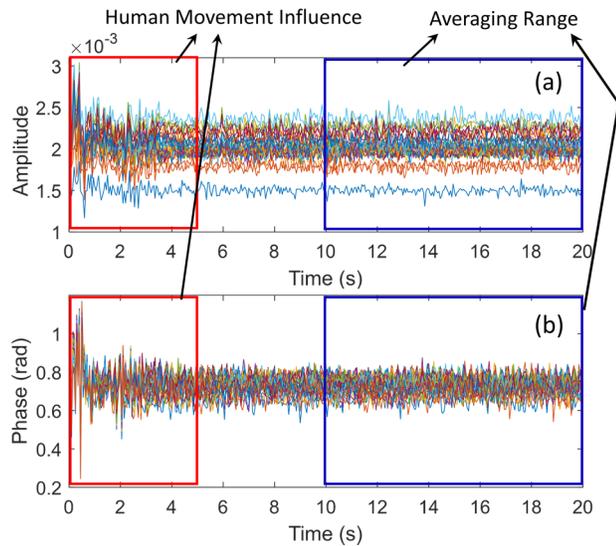

Fig. 7. CSI data of all recorded subcarriers at Rx2 after preprocessing in time dimension. (a) Amplitude. (b) Adjusted phase

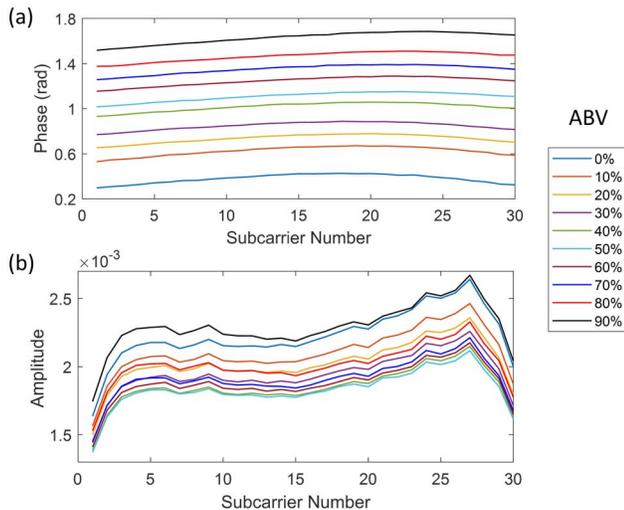

Fig. 8. Time-averaged CSI data of ethanol/water mixtures at Rx2 with respect to subcarrier. (a) Amplitude. (b) Adjusted phase.

### C. Dielectric Characterization using Theoretical Model

In the calibration stage as shown in Fig. 3, the measured dielectric properties at 5.32 GHz and the CSI data of the 16th subcarrier which is adjacent to the center frequency are utilized. The Levenberg-Marquardt method is applied to fit the complex-valued equation [48]. Then, using the estimated coefficients, the dielectric properties are calculated according to the workflow in the estimation stage. The relative error $\delta_{\varepsilon_r}$ and $\delta_\sigma$ are used to assess the accuracy:

$$\delta_{\varepsilon_r} = |\varepsilon_r - \hat{\varepsilon}_r| / \varepsilon_r \tag{22}$$

$$\delta_\sigma = |\sigma - \hat{\sigma}| / \sigma \tag{23}$$

where $\varepsilon_r$ and $\sigma$ are the dielectric properties measured by standard method. $\hat{\varepsilon}_r$ and $\hat{\sigma}$ are the corresponding estimated values using the proposed WiEps system.

The WiEps is first calibrated using the known ethanol/water mixtures. Then, the calibrated system is used to measure the unknown materials which are the Baijiu liquors. The estimated results and the comparison with standard method are shown in Table I. It can be observed that the relative errors of dielectric constant are lower than 8% and the average error across 12 materials is approximately 4.3%. This demonstrates that the proposed model effectively characterizes the dielectric constant using the CSI data. While the corresponding errors of conductivity are within 15% and the average error across 12 materials is about 7.7%, which is relatively larger compared with that of dielectric constant. These results indicate that the proposed WiEps can be used for the measurement of dielectric properties. Especially, the quantitative quality check of liquors is possible with WiEps. The ABV can be estimated using a pre-prepared table which lists different ABVs and the corresponding dielectric properties.

TABLE I
ESTIMATION RESULTS USING DATA FROM 16TH SUBCARRIER

| ABV | $\hat{\varepsilon}_r$ | $\varepsilon_r$ | $\delta_{\varepsilon_r}$ (%) | $\hat{\sigma}$ (S/m) | $\sigma$ (S/m) | $\delta_\sigma$ (%) |
|---|---|---|---|---|---|---|
| 0 % | 77.92 | 73.38 | 6.2 | 7.05 | 6.41 | 9.9 |
| 10 % | 57.17 | 57.12 | 0.1 | 7.44 | 8.33 | 10.5 |
| 20 % | 48.53 | 50.89 | 4.6 | 7.74 | 8.64 | 11.1 |
| 30 % | 39.90 | 40.64 | 1.8 | 7.86 | 8.57 | 8.1 |
| 40 % | 28.49 | 30.66 | 7.1 | 7.57 | 7.71 | 0.4 |
| 50 % | 23.46 | 24.74 | 5.2 | 7.12 | 6.82 | 3.3 |
| 60 % | 17.57 | 18.48 | 4.9 | 5.75 | 5.54 | 3.8 |
| 70 % | 13.42 | 13.72 | 2.2 | 4.78 | 4.32 | 12.5 |
| 80 % | 10.47 | 9.93 | 5.5 | 3.62 | 3.15 | 14.9 |
| 90 % | 7.36 | 6.85 | 7.5 | 1.74 | 2.02 | 15.2 |
| Baijiu 46% | 28.52 | 27.76 | 2.7 | 7.37 | 7.29 | 1.0 |
| Baijiu 56% | 22.23 | 21.33 | 4.2 | 6.25 | 6.13 | 2.0 |

### D. Estimation Results with Other Subcarriers

The bandwidth of the WiFi signal is 20 MHz and the frequency range of the operating channel is from 5310 MHz to 5330 MHz. Within the operating frequency range, the dielectric properties of the ethanol/water solutions are almost the same. If the values at the center frequency are chosen as the baseline, the variation range for $\varepsilon_r$ is [-0.17%, 0.18%] and that for $\sigma$ is [-0.44%, 0.51%]. Therefore, the dielectric properties can be regarded as constant within the bandwidth range.

The bandwidth is divided into 64 subcarriers with a frequency interval of 312.5 kHz and only 56 of them are used to transmit data and pilots. The recorded CSI data include a subgroup of 30 subcarriers from No. -28 to No. 28 [45]. The CSI data from other subcarriers are also applied to calibrate the WiEps system and to characterize the Baijiu liquors. The estimation errors are shown in Fig. 9. Each group of the bar plots represents the results for a certain material. The subcarrier is sorted in ascending order from left to right within the group. It can be observed that as for dielectric constant, the estimation errors are below 8% which is at the same level of the 16th subcarrier except for the 90% solution. The estimation errors with different subcarriers vary a lot in some materials such as the 90% solution and 46% Baijiu. The difference in error is larger than 5%. Meanwhile, some subcarriers perform better than the 16th one in some materials such as 30%, 70% solutions and both 46% and 56% Baijiu. Similar observation is also obtained for the conductivity error. Except for the data of 80% solution, the estimation errors are below 15% which is at the same level of the 16th subcarrier. The estimation results are quite similar for the 0%, 10%, 20%, 30% solutions. While those for 40%, 50%, 60% have larger variation. There is no significant evidence to imply which subcarrier performs best consistently. However, it is found that in most cases where the estimation error changes a lot, the error value is almost ascending or descending with respect to the subcarrier number. Therefore, the subcarrier which is adjacent to center frequency is considered to be a moderate choice for estimating the dielectric property.

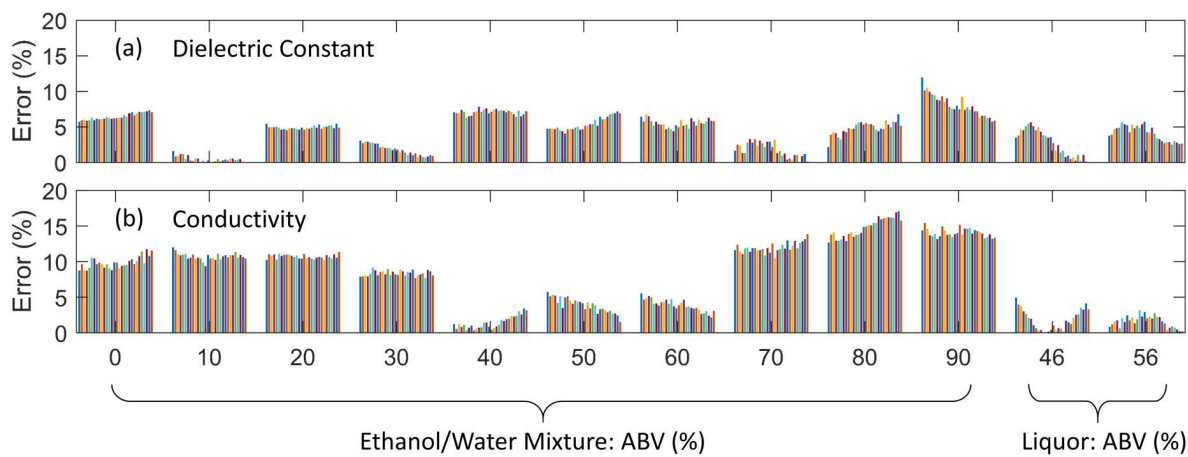

Fig. 9. The measurement error using other subcarriers for calibration and estimation. (a) Dielectric constant error. (b) Conductivity error.

### E. Validation with Other Liquids

In order to further validate the efficacy of the proposed WiEps, it is applied to more liquids. Besides, the experiment with no material inside the container (air) is also carried out. During the experiment, the distances between the object and Tx/Rx antenna are set as 10 cm and the distance between Tx and Rx antennas is 20 cm. The additional liquids include two other kinds of liquors (Jinro Soju, Grape Soju), saline with different concentrations (0.9%, 3.5%, 7%), and glucose/water mixtures (5%, 10%, 25%). For each material, the measurement is conducted 5 times and the order of the measurement is shuffled. Totally, 105 measurements are carried out with 21 materials (including air).

Figure 10(a) and 10(b) shows the amplitudes and phases of all measurement of the materials with respect to subcarrier number. By averaging the data from multiple measurements, the averaged values at 16th subcarrier are shown in Fig. 10(c). It can be observed that, the amplitude of air is the largest and those of the saline are small. This is because the conductivity of saline is relatively high. The WiEps is calibrated with known ethanol/water mixtures using the averaged data from multiple measurements. Then, the dielectric properties of other materials are estimated using the averaged data as well. The measurement results for all materials are shown in Table II. The dielectric properties of the air are estimated with absolute errors of 0.38 and 0.39 for $\varepsilon_r$ and $\sigma$. Since the actual dielectric properties for air is small, the relative errors can be a large number even when the absolute value is small. Therefore, when referring to the average relative error, the air is excluded. The average relative errors of dielectric constant and conductivity across all materials except air are about 4.0% and 8.9%. Compared with the experiment where only ethanol/water mixtures are tested, the accuracy is at the same level, demonstrating the feasibility and robustness of the proposed method.

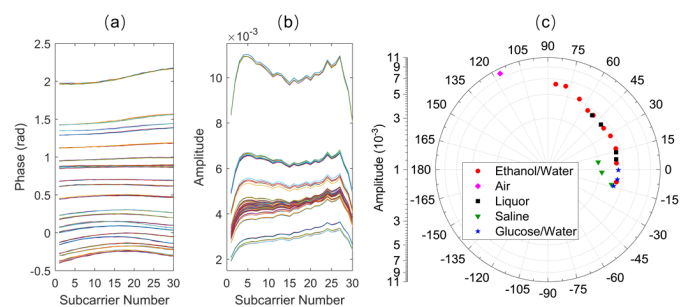

Fig. 10. CSI data of all measurements of the materials at Rx2 with respect to subcarrier when the distance between Tx and Rx is 20 cm. (a) Amplitude. (b) Adjusted phase. (c) Averaged phase/amplitude at 16th subcarrier from multiple measurements.

### F. Effect of Distance Between the Object and Antennas

The effect of the distance between the object and Tx/Rx antennas is investigated. Additional experiments are conducted when the distances between Tx and Rx antennas are 40 cm, 60 cm, and 100 cm. The distances between the object and Tx/Rx antenna are set equal. The measured amplitudes and phases of all experiments are shown in Fig. 11. It can be observed that the amplitude of the received signal is decreasing when the distance between antennas is increasing. This is caused by the effect of path loss of the microwave.

TABLE II
ESTIMATION RESULTS USING AVERAGED DATA AT 16TH SUBCARRIER FROM MULTIPLE MEASUREMENTS

| Material | | $\widehat{\varepsilon_r}$ | $\varepsilon_r$ | $\delta_{\varepsilon_r}$ (%) | $\widehat{\sigma}$ (S/m) | $\sigma$ (S/m) | $\delta_\sigma$ (%) |
|---|---|---|---|---|---|---|---|
| Ethanol/ Water Mixture | 0 % | 72.56 | 73.38 | 1.1 | 7.22 | 6.41 | 12.5 |
| | 10 % | 59.61 | 57.12 | 4.3 | 8.00 | 8.33 | 4.0 |
| | 20 % | 49.36 | 50.89 | 3.0 | 7.63 | 8.64 | 11.6 |
| | 30 % | 40.04 | 40.64 | 1.5 | 7.73 | 8.57 | 9.9 |
| | 40 % | 33.06 | 30.66 | 7.8 | 8.10 | 7.71 | 5.0 |
| | 50 % | 22.42 | 24.74 | 9.4 | 7.19 | 6.82 | 5.3 |
| | 60 % | 18.20 | 18.48 | 1.5 | 6.67 | 5.54 | 20.4 |
| | 70 % | 13.69 | 13.72 | 0.2 | 4.60 | 4.32 | 6.6 |
| | 80 % | 9.81 | 9.93 | 1.2 | 2.47 | 3.15 | 21.6 |
| | 90 % | 7.28 | 6.85 | 6.3 | 2.06 | 2.02 | 2.0 |
| Liquor | Baijiu 46% | 29.97 | 27.76 | 8.0 | 7.96 | 7.29 | 9.2 |
| | Baijiu 56% | 21.69 | 21.33 | 1.7 | 7.34 | 6.13 | 19.8 |
| | Grape Soju | 56.81 | 51.17 | 11.0 | 8.04 | 8.17 | 1.6 |
| | Jinro Soju | 51.97 | 50.01 | 3.9 | 7.80 | 8.48 | 8.0 |
| Saline | 0.9 % | 70.48 | 70.87 | 0.6 | 8.06 | 7.66 | 5.3 |
| | 3.5 % | 69.21 | 64.93 | 6.6 | 12.28 | 10.60 | 15.9 |
| | 7 % | 59.74 | 58.39 | 2.3 | 13.89 | 13.84 | 0.4 |
| Glucose/ Water Mixture | 5 % | 71.10 | 70.38 | 1.0 | 7.35 | 6.73 | 9.3 |
| | 10 % | 70.61 | 67.63 | 4.4 | 7.11 | 6.75 | 5.4 |
| | 25 % | 64.22 | 62.25 | 3.2 | 7.34 | 7.10 | 3.5 |
| Air | - | 1.38 | 1 | 37.9 | 0.39 | 0 | Inf |

Process the measured data with the same procedure as that in Section V.E. The average errors of dielectric constant when the distances are 40 cm, 60 cm, and 100 cm are about 9.3%, 10.2%, and 6.1%, respectively. The corresponding average errors of conductivity are 18.7%, 14.5%, and 13.8%, respectively. The accuracies are worse than the case when the distance is 20 cm. It can be noted that the results for 100 cm are better than those of 40 cm and 60 cm. The reason is considered to be the effect of multipath. When the distances are 40 cm and 60 cm, the phase spans are only 40 and 78 degrees, respectively. For comparison, those of 20 cm and 100 cm are 130 and 203 degrees, respectively. When the multipath signals are strong, the change in LOS component which caused by the materials have less influence on the total received signal. Therefore, the phase span is smaller. Since the data are converted from analog signals, the accuracy is degraded by the quantum error and small values may not be precisely sampled. The data accuracy is worse if the value range is small. In the proposed theoretical model, the effect of multipath signals is included. However, if the data accuracy is low, the performance of the model will be impaired. Thus, the performance when the distances are 40 cm and 60 cm are worse than that of 100 cm. Comparing the results for the cases of 20 cm and 100 cm, it can be found that the smaller distance has better estimation accuracy because the amplitude of the received signal is larger and the data accuracy can be improved. Therefore, to get an optimal performance of

WiEps with commodity WiFi device, the distance between object and the antennas should be configured relatively small. Meanwhile, weaker multipath signals with the configuration can help improve the measurement accuracy.

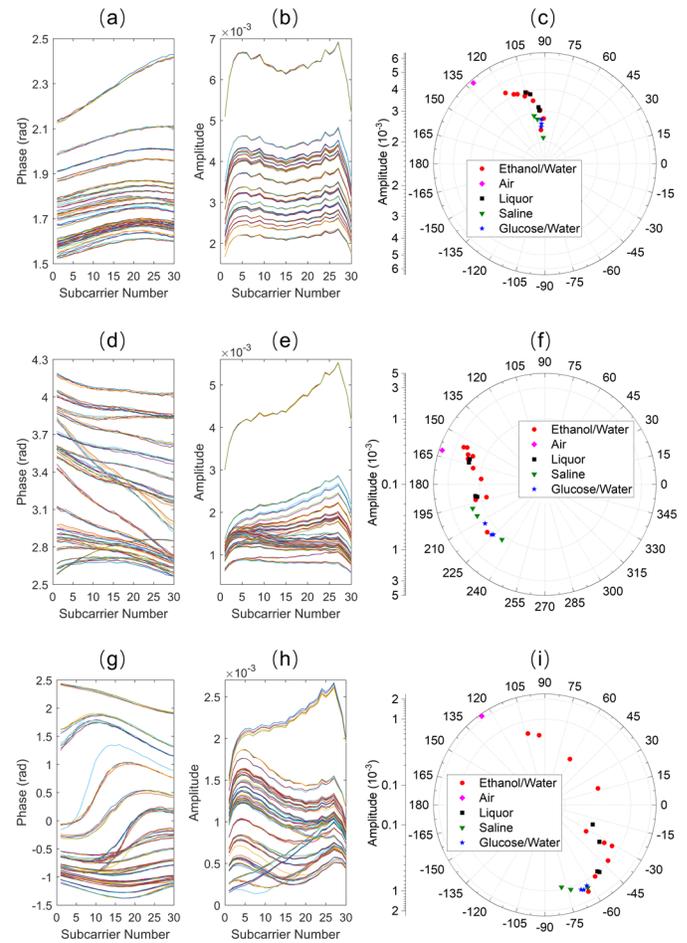

Fig. 11. Amplitude, adjusted phase, and averaged phase/amplitude at 16th subcarrier from multiple measurements of CSI data for all measurements when the distances between Tx and Rx are (a)~(c) 40 cm, (d)~(e) 60 cm, and (g)~(i) 100 cm, respectively.

## VI. CONCLUSION

A dielectric property measurement system, WiEps, is developed which is composed of the commodity WiFi devices. The proposed system utilizes the transmission features of the microwave caused by dielectric property change. The transmitted WiFi signal through the material is theoretically analyzed and it is utilized to quantitatively measure the dielectric properties of the materials. A theoretical model is proposed to characterize the wave propagation behavior through the material in the WiEps system and to quantitatively describe the relationship between the material dielectric property and CSI data. The system is first applied to measure the ethanol/water mixtures. Then, additional liquids are tested for further verification. In the experiment, WiEps is deployed under the indoor condition where the multipath effect exists. Using the WiEps, the average errors across all materials are 4.0% and 8.9% for dielectric constant and conductivity, respectively. These results demonstrate the efficacy of the proposed WiEps system for dielectric property measurement.

By quantitatively estimating the dielectric properties of the material, this method can be applied to food engineering such as quality check of liquors and identification of liquids. Besides, this method is also promising to be applied in security check, water contamination detection, and monitoring the dielectric property of material during manufacturing process. More applications will be considered and investigated in the future work.

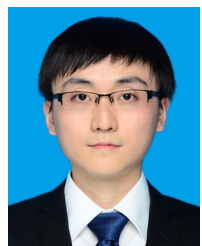
**Hang Song** received the B.S. and M.S. degrees in electronic science and technology from Tianjin University, Tianjin, China, in 2012 and 2015, respectively. He received the Ph.D. degree from Hiroshima University, Hiroshima, Japan in 2018. He was a visiting researcher at Research Institute for Nanodevice and Bio Systems (RNBS), Hiroshima University. He is currently an assistant professor with Tianjin key laboratory of imaging and sensing microelectronic technology, school of microelectronics, Tianjin University, China.

His research interests are wireless sensing, microwave imaging, signal processing, complex permittivity measurement for biomedical engineering, microwave detection system development, and antenna design.

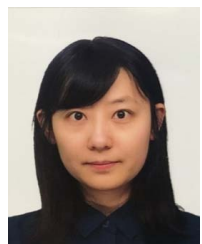
**Bo Wei** received the B.E. and M.E. degrees from Tianjin University, Tianjin, China, in 2012 and 2015, respectively. She received the Ph.D. degree from Waseda University, Tokyo, Japan in 2019. She is currently an assistant professor with the Graduate School of Fundamental Science and Engineering, Waseda University.

Her research interests include wireless communication, machine learning, adaptive video transmission, computer networking, and Internet of Things. She is a member of the IEEE and IEICE.

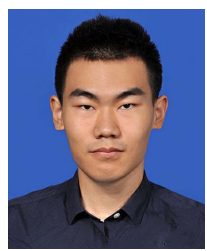
**Qun Yu** received the B.S. degree in Electronic Science and technology from Tianjin University, Tianjin, China in 2018. He is currently a Master student in microelectronics and solid electronics at Tianjin University. His research interests including microwave detection system and machine learning.

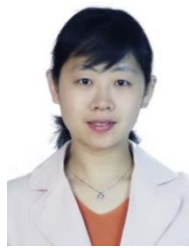
**Xia Xiao** (M'01) received the B.S degree in physics and the M.S. degree in condensed physics from Tianjin Normal University, Tianjin, China, in 1993 and 1996, respectively, and the Ph.D. degree in electronic and information technology from the Technical University of Chemnitz (TU Chemnitz), Chemnitz, Germany, in 2002.

From 2002–2003, she contributed to the "MIRAI Project" at the National Institute of Industrial Science and Technology (AIST), Tokyo, Japan, where she worked in ULSI low-k/Cu interconnect technology as a Key Researcher. In 2003, she joined the School of Electronic Information Engineering, Tianjin University, Tianjin, China, where she is currently a Professor. From 2006–2007, she was a Visiting Professor at Hiroshima University, Hiroshima, Japan, where she worked in developing algorithms for UWB imaging for early breast cancer detection. Her research interests include advanced algorithms for early breast cancer detection by UWB and non-destructive characterization of film properties by surface acoustic waves (SAWs).

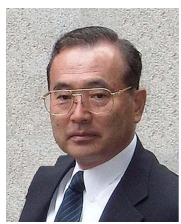
**Takamaro Kikkawa** (S'74–M'76–SM'01–F'10) received the B.S. and M.S. degrees in electronic engineering from Shizuoka University, Shizuoka, Japan, in 1974 and 1976, respectively, and the Ph.D. degree in electronic system from the Tokyo Institute of Technology, Tokyo, Japan, in 1994. In 1976, he joined the NEC Corporation, Tokyo, Japan, where he conducted research and development on interconnect technologies for large scale integrated circuits and dynamic random access memories. From 1983 to 1984, he was the Visiting Scientist at the Massachusetts Institute of Technology, Cambridge, MA, USA, where he conducted research on SOI transistors. In 1998, he joined the faculty of Hiroshima University, Hiroshima, Japan, where he is Professor of the Graduate School of Advanced Sciences of Matter and Director of the Research Institute for Nanodevice and Bio Systems. He is also Councilor of Hiroshima University. From 2001 to 2008, he was appointed the Senior Research Scientist at the National Institute of Advanced Industrial Science and Technology, Tsukuba, Japan, and the Group Leader of Low-k/Cu Interconnect Technology of Japan's "MIRAI Project." His research interests include wireless and wired interconnect technologies, impulse-radio-CMOS transceiver circuits with on-chip antennas, and impulse-radar-based CMOS breast cancer detection systems.

Dr. Kikkawa is a Fellow of Japan Society of Applied Physics.